\documentclass[aps,prc,twocolumn,groupedaddress,floatfix,showpacs,twoside]{revtex4}
\usepackage{graphicx}
\usepackage{dcolumn}
\usepackage{bm}
\begin{document}

\setcounter{page}{1}
\pagestyle{myheadings} \oddsidemargin
4mm\evensidemargin -4mm

\title{\bf Evidence for a Novel Reaction Mechanism of a Prompt Shock-Induced Fission Following the Fusion of $^{78}$Kr and $^{40}$Ca Nuclei at E/A =10 MeV\\}

\author{E.~Henry,$^1$ J.~T{\~o}ke,$^1$ S.~Nyibule,$^2$ M.~Quinlan,$^1$ W.U.~Schr{\"o}der,$^{1,2}$
G.~Ademard,$^3$ F.~Amorini,$^4$ L.~Auditore,$^{5,6}4$ C.~Beck,$^7$ I.~Berceanu,$^8$ E.~Bonnet,$^3$ B.~Borderie,$^9$ G. Cardella,$^{10}$ A.~Chbihi,$^3$ M.~Colonna,$^4$
E.~De~Filippo,$^{10}$ A.~D'Onofrio,$^{11,12}$ J.D.~Frankland,$^3$ E.~Geraci,$^{13,10}$ E.~La~Guidara,$^{10,13}$ M.~La~Commara,$^{14,11}$ G.~Lanzalone,$^{15,4}$
P.~Lautesse,$^{16}$ D.~Lebhertz,$^3$ N.~Le~Neindre,$^{17}$ I.~Lombardo,$^{18,4}$ D.~Loria,$^{5,6}$ K.~Mazurek,$^3$ A.~Pagano,$^{10}$ M.~Papa,$^{10}$ E.~Piasecki,$^{19}$ S.~Pirrone,$^{10}$
G.~Politi,$^{18,10}$ F.~Porto,$^{18,4}$ F.~Rizzo,$^{18,4}$ E.~Rosato,$^{14,11}$ P.~Rusotto,$^{18,4}$ G.~Spadaccini,$^{14,11}$ A.~Trifir{\'o},$^{5,6}$ M.~Trimarchi,$^{5,6}$ G.~Verde,$^{10}$ M.~Vigilante,$^{14,11}$ and J.P.~Wieleczko$^3$}

\address{$^{1}$ Dept. of Chemistry - University of Rochester, USA\\
        $^{2}$ Dept. of Physics - University of Rochester, USA\\
        $^{3}$ GANIL Caen, France\\
        $^{4}$ INFN Laboratori Nazionali del Sud, Italy\\
        $^{5}$ Dipartamento di Fisica Universit{\`a} di Messina, Italy\\
        $^{6}$ INFN Gruppo Collegiato di Messina, Italy\\
        $^7$ IN2P3/CNRS - IPHC and Universit{\'e} de Strasbourg, France\\
        $^8$ IPNE Bucharest, Romania\\
        $^9$ IN2P3 - IPN Orsay, France\\
        $^{10}$ INFN Sezione di Catania, Italy\\
        $^{11}$ INFN Sezione di Napoli, Italy\\
        $^{12}$ Dipartimento di Scienze Ambientali Seconda Universit{\`a} di Napoli, Caserta, Italy\\
        $^{13}$ Centro Siciliano Fisica Nucleare e Struttura della Materia, Catania, Italy\\
        $^{14}$ Dipartamento di Scienze Fisiche Universit{\`a} Federico II Napoli, Italy\\
        $^{15}$ Universit{\`a} Kore, Enna, Italy\\
        $^{16}$ IN2P3 - Lyon, France\\
        $^{17}$ IN2P3 - LPC Caen, France\\
        $^{18}$ Dipartamento di Fisica e Astronomia Universit{\`a} di Catania, Italy\\
       $^{19}$ Warsaw University, Poland\\
}
\begin{abstract}
An analysis of experimental data from the inverse-kinematics ISODEC experiment on $^{78}$Kr+$^{40}$Ca reaction at a bombarding energy of 10 AMeV has revealed signatures of a hitherto unknown reaction mechanism, intermediate between the classical damped binary collisions and fusion-fission, but also substantially different from what is being termed in the literature as fast fission or quasi fission. These signatures point to a scenario where the system fuses transiently while virtually equilibrating mass asymmetry and energy and, yet, keeping part of the energy stored in a collective shock-imparted and, possibly, angular momentum bearing form of excitation. Subsequently the system fissions dynamically along the collision or shock axis with the emerging fragments featuring a broad mass spectrum centered around symmetric fission, relative velocities somewhat higher along the fission axis than in transverse direction, and virtually no intrinsic spin. The class of mass-asymmetric fission events shows a distinct preference for the more massive fragments to proceed along the beam direction, a characteristic reminiscent of that reported earlier for dynamic fragmentation of projectile-like fragments alone and pointing to the memory of the initial mass and velocity distribution.
\end{abstract}

\maketitle

Studies of heavy-ion reaction dynamics in the energy domain moderately in excess of the Coulomb barrier have led, already in their early days of nineteen seventies and eighties, \cite{wilcz_1973} to the emergence of a general scenario of dissipative collisions \cite{schroeder_huizenga} at peripheral and mid-peripheral collisions, then transiting over to fast- or quasi-fission at more central collisions \cite{heusch_qf,lebrun_qf,borderie_qf,gregoire_qf,toke_qf_PL,toke_qf_NP,shen_1987} and ultimately leading to fusion at even more central collisions. Subsequently, it was found \cite{baldwin,djerroud} that such a ``gentle'' scenario is still largely valid in the domain of intermediate or Fermi bombarding energies, however, with the role of dynamically-induced decay modes progressively increasing with increasing bombarding energy. Among the latter are pre-equilibrium emission of nucleons \cite{bondorf_1976,randrup_vandenbosch,wile_1989}, what has been termed ``neck'' emission of intermediate-mass fragments \cite{montoya,toke_neck}, and dynamic fragmentation of projectile-like fragments \cite{de_souza_2002,de_souza_2010}. It is important to note in the context of the present study that while there are plausible theories explaining the phenomenology of dissipative collisions, \cite{randrup_NEM} fusion, pre-equilibrium particle emission, \cite{bondorf_1976,randrup_vandenbosch} and, perhaps also the quasi-fission, \cite{gregoire_qf} there still are no such to describe satisfactorily the ``neck'' emission of fragments and the dynamic PLF fragmentation. The latter tend to point to a possible role of relevant collective degrees of freedom, such that would readily couple to the fragmentation channels, being excited in the course of more violent collisions.

\begin{figure}
\includegraphics [scale = .5]{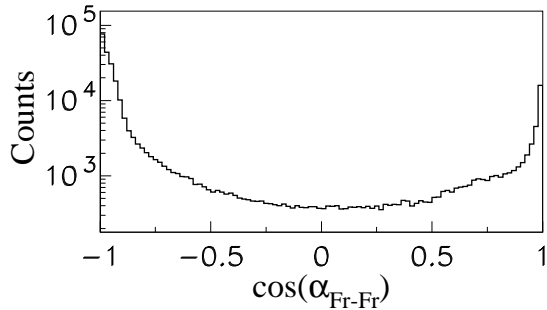}
\caption{Distribution of the cosine of the folding angle $\alpha$ between the C.M. velocities of two fragments with $Z>3$.}
\label{fig_1_collinear}
\end{figure}

The present study reports a possible discovery of yet another class of collisions with clear signatures of a dynamic process, pointing this time, perhaps,  more clearly to the role of collective excitations in the prompt decay of the system into binary exit channels. Importantly, the process is characterized by significant cross section and is here named ``shock-induced fission following fusion'' to reflect the nature of the observed robust signatures. A fuller account of all the effort taken to ascertain that these signatures are not an artifact of the deficiencies in the detection setup and/or analysis routines will be given elsewhere. Here, the most essential results are discussed along with their significance and an interpretation.

\begin{figure}
\includegraphics [scale = .5]{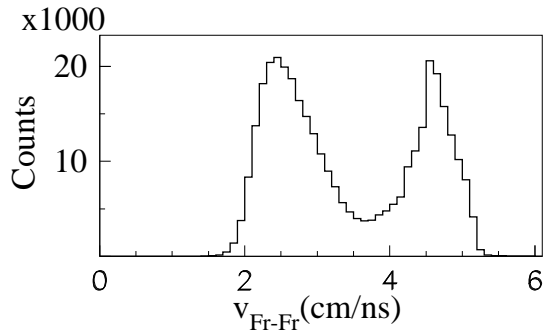}
\caption{Distribution of relative velocities of the fragments satisfying the collinearity condition of
$-1.0<cos(\alpha)<-0.7$.}
\label{fig_2_relvel}
\end{figure}

The experimental data used in the present study constitute a fragment of a large set of data  on $^{78}$Kr+$^{40}$Ca and $^{86}$Kr+$^{48}$Ca reactions at 10 AMeV, collected in the ISODEC experiment. The experiment was performed at the INFN-Laboratori Nazionali del Sud (LNS) and utilized the highly efficient CHIMERA detector. \cite{chimera} Data from only the first of these two systems were used in the present study. While all energy calibrations and fragment identifications were performed earlier without regard to the present analysis, of particular importance for the present study is the $A,Z$ identification. This relied on energy measured in the silicon detectors alone and time-of-flight measurement with non-trivial time-offset evaluation. The latter relied on the recursive methodology developed earlier at LNS \cite{t_zero} that accounted for the pulse rise-time dependence on the fragment specie and energy. The measured energies were subsequently corrected for $A$- and $Z$- dependent pulse-hight defects according to the well-known Moulton formula \cite{moulton}. All energies were subsequently corrected for the losses in the $1-mg/cm^2$ $^{40}$Ca target, as was the beam energy. These corrections were found to be of significance in the analysis of the center-of-mass velocity of the decaying system. The efficiency of the CHIMERA setup for the particular class of events of interest here was evaluated by first producing an azimuthally isotropic (in angle $\phi$) replica of the actual events and then subjecting it to the filter representing the established earlier geometric and electronic (energies and particle IDs) coverage of the CHIMERA detector in the present experiment. This ``azimuthal'' efficiency was found to be $\epsilon_{\phi} \approx 0.4$. Obviously, a setup such as CHIMERA generically lacks coverage of small angles. In this respect, further simulations using event generators in conjunction with the filter showed that the efficiency with respect to the C.M. angle $\theta_{CM}$ was practically 1 for angles in the range of $25^{\circ} < \Theta_{Fr} < 155^{\circ}$ and zero for angles outside the range of $15^{\circ} < \Theta_{Fr} < 165^{\circ}$. The efficiency with respect to the energy and particle ID was also practically 1 (for the class of events considered) in view of the substantial velocity boost due to the inverse kinematics.

The experimental observations are illustrated in the set of figures 1 - 6. The following brief ``roadmap'' should help one to better understand the rationale behind and the significance of these figures. One notes first that the analysis concentrated on a subset of events with two or more fragments present with atomic number $Z>3$ and aimed originally at identifying the yield of fission- or quasi-fission-like processes involving the system as a whole or the projectile-like ($^{78}$Kr) entities (PLF) alone. As it became obvious that the latter kind of events (fission of PLFs) is virtually absent, gating condition were selected to isolate the yields of fusion-fission like processes. These included conditions on the fragment collinearity in the center-of-mass system and, to eliminate the contribution from dissipative collisions, conditions on the relative velocity of fragments. This led then to a narrower subset of events with well defined total mass number $A_{tot}$ peaking at approximately 106 mass units and with a FWHM of approx 17 units, as determined based on the left slope of the distribution. On the high-$A_{tot}$ side, there is a well-pronounced contribution from events with incorrectly reconstructed velocities (TOF) and a cut-off was set at $A_{tot}$ = 118 to avoid contamination of various velocity-based plots by this subset of partially corrupted events. This cut-off affects approximately $30\%$  of events classified as binary fission. Subsequently, it was determined that the ``working'' set of events with $60 < A_{tot} < 118$ features gross characteristics inconsistent with known phenomenologies, but pointing consistently to a peculiar dynamics-driven scenario as discussed below. Importantly, the partially corrupted events with reconstructed $A_{tot}>118$ reveal the same gross characteristics, except for an approximately 0.1 - 0.25 cm/ns deficit in velocities. Accordingly these events were included in the evaluation of the cross section of interest.

Fig.~\ref{fig_1_collinear} illustrates the inclusive collinearity of two largest fragments with $Z>3$
measured  as a cosine of the angle $\alpha$ between the center-of-mass velocities of the fragments. The narrow well-defined peak at $cos(\alpha)=-1$ corresponds to fragments proceeding in opposite directions in the C.M. system as should be the case for fission-like processes of interest. Not surprisingly, there is a yield present representing background of various origin, but the overall quality of the collinearity spectrum well justifies setting a ``liberal'' gate in the range from -1.0 to -0.7 to select the fusion-fission like events for the further analysis. It is worth noting that Fig.~\ref{fig_1_collinear} indicates that the reconstructed center-of-mass velocity of the two fragments coincides on average well with such true velocity pointing to a quite satisfactory, for the present purpose, quality of the energy and (non-trivial) mass calibration. One notes also that a gate that is too narrow has the potential of biasing the selection against fission in transverse directions with respect to the beam axis, as the reconstructed fragment center of mass velocity shows more spread in the longitudinal as compared to the transverse direction.

\begin{figure}
\includegraphics [scale = .5]{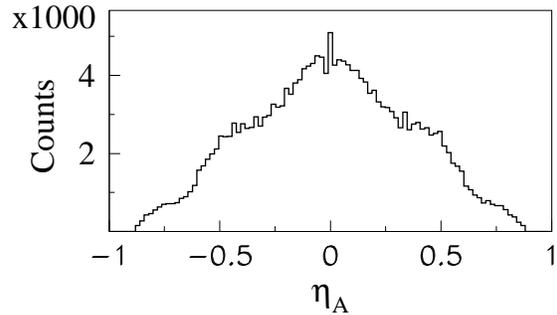}
\caption{Distribution of the mass asymmetry $\eta_A=(A_1-A_2)/(A_1+A_2)$ of fission-like fragments.}
\label{fig_3_aasy}
\end{figure}

Fig.~\ref{fig_2_relvel} illustrates the spectrum of relative velocities of fragments, $v_{Fr-Fr}$ with fission-like fragments being represented by the peak centered approximately around the Viola velocity and well separated from the peak corresponding to elastic, quasi-elastic, and dissipative collision events. For the purpose of the further study a gate was selected in the range from $v_{Fr-Fr}$ = 1.5 to 3.5 cm/ns. One must keep in mind that such a gate would naturally include fully-damped dissipative collision events as well as quasi-fission events and additional criteria need to be employed to assess the role of the latter. The criteria used in this study involve the characteristics of the fragment mass distribution, fragment angular distribution, and fragment spin distribution, and they all speak against significant contribution to the discussed yields from these two well-known processes, as well as from classical fusion-fission reactions.

\begin{figure}
\includegraphics [scale = .5]{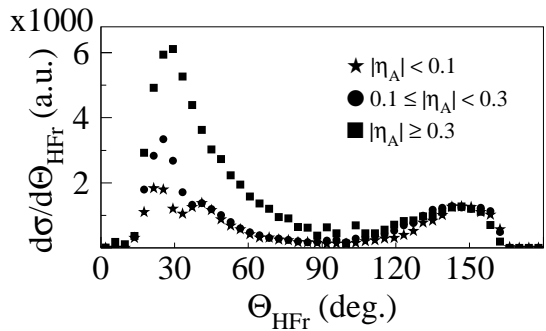}
\caption{Angular distributions of larger (of two) fragments in the center-of-mass system for three bins in mass asymmetry parameter $\eta_A$ and normalized to yields at backward angles.}
\label{fig_4_angdis}
\end{figure}

The reconstructed distribution of the sum of the fragment mass numbers for the class of events satisfying the two gating conditions - on collinearity and relative fragment velocity, shows a single peak with maximum around mass number of $A_{tot}$ = 106, consistent with complete fusion. Consistent with fusion-fission is also the mass asymmetry spectrum shown in Fig.~\ref{fig_3_aasy} featuring a peak centered around zero asymmetry. The mass asymmetry is here defined as $\eta_A=(A_1-A_2)/(A_1+A_2)$. As seen in this figure, there is little if any memory of the initial mass asymmetry of $\eta_A=0.322$ discernible, with irregularities readily attributable to experimental uncertainties in the non-trivial fragment mass reconstruction.
\begin{figure}
\includegraphics [scale = .5]{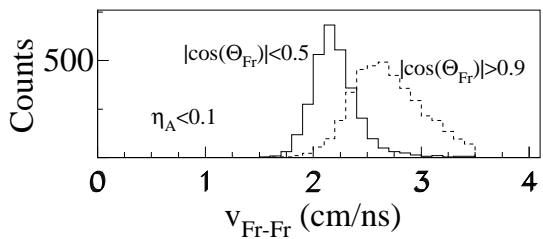}
\caption{Relative velocity spectra of fragments from mass-symmetric splits emerging in the beam (dashes) and in the transverse direction (solid line).}
\label{fig_5_relvels}
\end{figure}

\begin{figure}
\includegraphics [scale = .5]{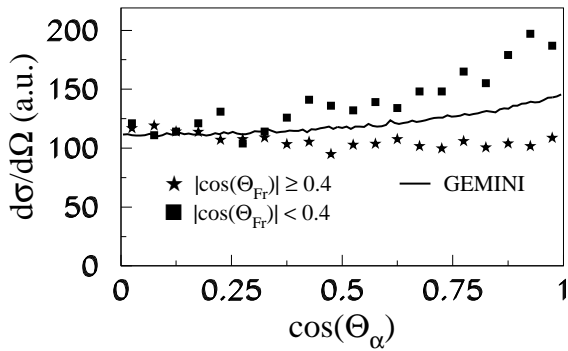}
\caption{Angular distribution of alpha particles emitted from heavier fragments in the forward hemisphere (away from the lighter fragment) relative to the fragmentation axis aligned with the beam axis (solid stars) and perpendicular to the beam axis (solid boxes). The solid line illustrates predictions by the code Gemini for emission from excited $^{58}$Ni fragments. (See text)}
\label{fig_6_spin}
\end{figure}

Figures 4 through 6 illustrate what is the crux of the present discovery - the irrotational character of the fragment angular distribution and of the fragmentation process as a whole. Fig.~\ref{fig_4_angdis} illustrates the center-of-mass angular distributions $d\sigma/d\Theta_{HFr}$ of the larger of the two fragments for three bins in mass asymmetry parameter $\eta_A$. In this representation, the decay of spinning compound systems would be in the limit uniform, but practically would always show maximum at $90^{\circ}$. Here, the angular distribution shows very strong forward-backward peaking with a distinct preference for the larger fragment to proceed in the beam direction. This asymmetry increases with the increasing mass-asymmetry $\eta_A$, as illustrated by the sequence seen in Fig~\ref{fig_4_angdis}. One notes that in classical quasi-fission, angular distributions were found to be consistent with those for fission \cite{toke_qf_PL,toke_qf_NP}, even as they were at times significantly more forward-backward peaked than expected on the grounds of the rotating liquid drop model. \cite{rld} The strong forward-backward peaking of the angular distribution is indicative of a dynamical irrotational process such as, e.g., caused by shock-induced intermediate storage of a part of kinetic energy in a collective vibrational mode and the subsequent reuse of this energy in the system fragmentation. Such a mode has a potential to ``memorize'' not only the direction of the shock but also the collective angular momentum involved, with this memory then leaving distinct experimental signatures, as displayed in Figs.~\ref{fig_5_relvels} and \ref{fig_6_spin}.

Fig.~\ref{fig_5_relvels} illustrates relative velocities of fragments emerging along the beam direction and in transverse direction and the fact that these velocities are higher in the former case. This trend is consistent with the supposition that the fragmentation process relies on an ``extra push'' from an intermediate collective energy reserve. Fig.~\ref{fig_6_spin} further demonstrates that the fragments emerging along the beam axis are virtually spinless, again consistent with an irrotational fragmentation. Specifically, this figure shows that the angular distribution of alpha particles emitted from heavier fragments moving along the beam direction is isotropic (depicted by stars), as expected for the case of emission from spinless entities. In contrast, the angular distribution for alpha particles emitted from fragments moving in transverse direction (depicted by boxed) shows a distinct anisotropy consistent with non-zero fragment spins, such as expected both, for fusion-fission and for quasi-fission, and thus indicative of the relevant part of the fragment yield being attributable to these well-known processes. This figure displays also the prediction by the code GEMINI \cite{gemini} for the emission of $\alpha$-particles from $^{58}$Ni fragments with spins in the (sticking limit) range of $I=0 - 12$ (depicted by solid line) and excitation energy of 2 AMeV, typical for fission fragments in this case.

In summary, a novel dynamical process termed shock-induced fission (SIF) has been discovered in heavy-ion collisions at energies of a few AMeV above the interaction barrier. In a classification scheme according to the impact parameter, this mechanism appears to be active in most central collisions where it leads to a complete, albeit transient fusion of the system with a strong memory of the initial impact direction and also a weak memory of the initial joint mass and velocity distribution. Given the evaluated cross section of the shock-induced fission process of $\sigma_{SIF} \approx$ 150 mb, the observed yield accounts for the range of initial angular momenta of $l \approx 0 - 40$ under the assumption of the centrality of collisions. While it is conceptually difficult to place the observed process between the fully damped dissipative collision and fusion, in that range the observed cross section would imply an angular momentum window of $l \approx 75-85$.

Following the fusion, the system re-separates promptly and predominantly along the direction well aligned with the beam axis into binary fission-like channel with a symmetric fragment mass distribution. The fragments appear not to have acquired any intrinsic angular momentum in the process, reinforcing the view of the process as an irrotational and shock-induced one. Consistent with such a view is the fact that it has not been observed at lower bombarding energies. One may speculate that the system re-separates in fact at small but always positive angles such that the initial angular momentum is found almost entirely in the relative angular momentum of the two fragments, but not in the fragments themselves. This could be experimentally verified in dedicated experiments covering the range of very small angles. Furthermore, in order to provide a more complete set of clues for the theoretical modeling of the observed process, it would be also of interest to check how the fragment mass distribution depends on the size of the system and, notably, if smaller systems would fragment mass-asymmetrically as expected for purely phase-space driven processes ({\sl vide} the concept of the Businaro-Gallone point). Also, of interest is a systematic exploration of the way the discovered mechanism sets in on the bombarding energy scale and then, possibly, fades away.

\begin{acknowledgments}
This work was supported by the U.S. Department of Energy grant No.
DE-FG02-88ER40414.
The authors wish to acknowledge valuable and stimulating discussions with L.G.~Sobotka and R.T.~de Souza.

\end{acknowledgments}


\end{document}